\documentstyle[12pt]{article}
\date{}

\begin{document}
\vspace*{-1.8cm}
\begin{flushright}
\flushright{\bf LAL/RT 00-15}\\
\vspace*{-0.2cm}
\flushright{December 2000}
\end{flushright}
\vskip 1. cm
\begin{center}
{\Large \bf Analytical Estimation of
the Beam-Beam\\Interaction Limited Dynamic Apertures \\and Lifetimes 
in e$^+$e$^-$ Circular Colliders}\\
\vspace{8mm}
{\large \bf J. Gao}\\
\vspace{3mm}
{\bf Laboratoire de L'Acc\'el\'erateur Lin\'eaire,}\\
IN2P3-CNRS et Universit\'e de Paris-Sud, BP 34, 91898 Orsay cedex, France\\
\end{center}
\vspace{6mm}

%
%
\baselineskip=14.5pt

\begin{abstract}
Physically speaking, the delta function like beam-beam nonlinear forces at interaction points (IPs) act as a sum of delta function nonlinear multipoles.
By applying the general theory established in ref. \cite{1},
in this paper we investigate analytically the beam-beam interaction limited
dynamic apertures and the corresponding beam lifetimes 
for both the round and the flat beams.
Relations between the beam-beam limited beam lifetimes and
the beam-beam tune shifts are established, which show clearly why
experimentally one has always a maximum beam-beam tune shift, $\xi_{y, max}$,
around 0.045 for e$^+$e$^-$ circular colliders, 
and why one can use round beams to double this value 
approximately.  
Comparisons with some machine parameters are given. Finally, we 
discuss the mechanism of the luminosity reduction due to a definite 
collision crossing angle.
\par
\end{abstract}
\section{Introduction}
Beam-beam interactions in circular colliders have many influences on 
the performance of the machines, 
and the most important effect is that 
beam-beam interactions contribute to the limitations on dynamic apertures
and beam lifetimes. 
Due to the importance of this subject, enormous efforts
have been made
to calculate incoherent and 
coherent beam-beam forces, to simulate beam-beam effects,
to find the difference between flat and round colliding
beams, and to establish analytical formulae to estimate
the maximum beam-beam tune shift \cite{9}-\cite{GG}. 
Physically speaking, the delta function like beam-beam nonlinear forces at interaction points (IPs) act as a sum of delta function nonlinear multipoles. In ref. \cite{1} we have established a general theory to 
study analytically in detail the delta function 
multipoles and their combined effects on the dynamic apertures
in circular storage rings, and  
in this paper we will apply these general analytical formulae 
to the case of beam-beam interactions and 
find the corresponding beam dynamic apertures and beam lifetimes. 
We will show quantitatively why there exists a maximum 
beam-beam tune shift, $\xi_{y,max}$, around 0.045 for flat beams in 
e$^+$e$^-$ circular colliders, and why this number can be almost doubled
for round colliding beams. Applications to some machine parameters are also 
given.  
In this
paper we will restrict ourselves to the discussion of
e$^+$e$^-$ circular colliders since the
treatment for the hadron colliders will be somewhat different and 
more difficult. 
Finally, we discuss beam-beam effects with a definite crossing angle. 
\par
\section{Beam-beam interactions}
For two head-on colliding bunches, the 
incoherent kick felt by each particle can be 
calculated as \cite{2a}:
\begin{equation}
\delta y'+i\delta x'=-{N_er_e\over \gamma_* }f(x,y,\sigma_x,\sigma_y)
\label{eq:111}
\end{equation}
where $x'$ and $y'$ are the horizontal and vertical 
slopes, $N_e$ is the particle population in the bunch, $r_e$ is the 
electron classical radius (2.818$\times 10^{-15}$ m), 
$\sigma_x$ and $\sigma_y$ are 
the standard deviations of the transverse charge density distribution
of the counter-rotating bunch at IP, $\gamma_*$ is the normalized particle's
energy, and $*$ denotes the test particle and the
bunch to which the test particle belongs.
When the bunch is Gaussian $f(x,y,\sigma_x,\sigma_y)$ can be expressed
by Basseti-Erskine formula \cite{Bast}: 
$$f(x,y,\sigma_x,\sigma_y)=\sqrt{2\pi \over \sigma_x^2-\sigma_y^2}\times$$
\begin{equation}
\left(
w\left({x+iy\over \sqrt{2(\sigma_x^2-\sigma_y^2)}}\right)
-\exp\left(-{x^2\over 2\sigma_x^2}-{y^2\over 2\sigma_y^2}\right)
w\left({{\sigma_y\over \sigma_x}x+i{\sigma_x\over \sigma_y}y\over \sqrt{2(\sigma_x^2-\sigma_y^2)}}\right)\right)
\label{eq:1111a}
\end{equation}
where $w$ is the complex error function expressed as 
\begin{equation}
w(z)=\exp(-z^2)(
1-{\rm erf}(-iz))
\label{eq:112a}
\end{equation} 
For the round beam ({\rm RB}) and the flat beam ({\rm FB}) cases 
one has the incoherent beam-beam kicks expressed as  
\cite{2}\cite{2ac}\cite{2a}:
\begin{equation}
\delta r'=-{2N_er_e\over \gamma_* r}\left(1-\exp\left(-{r^2\over 2\sigma^2}
\right)\right)\qquad ({\rm RB}, \sigma_x=\sigma_y=\sigma)
\label{eq:1}
\end{equation}
\begin{equation}
\delta x'=-{2\sqrt{2}N_er_e\over \gamma_* \sigma_x}\exp\left(-{x^2\over 
2\sigma_x^2}\right)\int_0^{x\over \sqrt{2}\sigma_x}\exp(u^2)du
\qquad ({\rm FB}, \sigma_x>>\sigma_y)
\label{eq:1x}
\end{equation}
\begin{equation}
\delta y'=-{\sqrt{2\pi}N_er_e\over \gamma_* \sigma_x}\exp\left(-{x^2\over 
2\sigma_x^2}\right){\rm erf}\left({y\over \sqrt{2}\sigma_y}\right)
\qquad ({\rm FB}, \sigma_x>>\sigma_y)
\label{eq:1a}
\end{equation}
where $r=\sqrt{x^2+y^2}$.
Now we want to calculate the average kick felt by the test particle
since the probability to find the transverse displacement of the test particle
is not constant (in fact, the probability function is the same as the charge
distribution of the bunch to which the test particle belongs in lepton machines due to synchrotron radiations).
In the following we assume that the transverse sizes for the two
colliding bunches at IP are exactly the same. 
For the round beam case after averaging one gets\cite{2}\cite{2aa}:
\begin{equation}
\delta \bar{r}'=-{2N_er_e\over \gamma_* \bar{r}}
\left(1-\exp\left(-{\bar{r}^2\over 4\sigma^2}
\right)\right)\qquad ({\rm RB})
\label{eq:1p}
\end{equation} 
Although this expression is the 
same as that of the coherent beam-beam kick for round beams, 
one should keep in mind 
that we are not finding coherent beam-beam kick originally, and the difference
will be obvious when we treat the vertical motion in the case of 
flat beams. 
For the flat beam case, we will treat the horizontal and 
vertical planes separately. As far as the horizontal kick is concerned, 
the horizontal kick depends only on one displacement variable just similar
to the round beam case, we will use its coherent form expressed as follows
\cite{2ac}\cite{2aa}:
\begin{equation}
\delta x'=-{2N_er_e\over \gamma_* \sigma_x}\exp\left(-{x^2\over 
4\sigma_x^2}\right)\int_0^{x\over 2\sigma_x}\exp(u^2)du
\qquad ({\rm FB})
\label{eq:1xx}
\end{equation}
where $\sigma_x$ in the incoherent formula in ref. \cite{2ac} has been
replaced by $\Sigma_x=\sqrt{2}\sigma_x$ (for two identical 
Gaussian colliding beams) according to Hirata theorem
demonstrated in the appendix A of ref. \cite{2aa}. 
As for the vertical kick, however,
one has to make an average over eq. \ref{eq:1a} 
with the horizontal
probability distribution function of the test particle, and one gets \cite{2a}:
\begin{equation}
\delta y'=-{\sqrt{2\pi}N_er_e\over \gamma_* \sigma_x}<\exp\left(-{x^2\over 
2\sigma_x^2}\right)>_x{\rm erf}\left({y\over \sqrt{2}\sigma_y}\right)
\qquad ({\rm FB})
\label{eq:111a}
\end{equation}
where $<>_x$ means the average over the horizontal
probability distribution function of the test particle, and for two
identical colliding Gaussian
beams $<>_x=1/\sqrt{2}$.
It is obvious that eq. \ref{eq:111a} is not the expression for the 
coherent beam-beam kick. The average over eqs. \ref{eq:1} and \ref{eq:1a}
is only a technical operation to simplify (or to make equivalence) 
a two dimensional problem 
to a one dimensional one. 
To study both round and flat beam cases, 
we expand $\delta  \bar{r}'$ at $x=0$ (for round beam we study only 
vertical plane since the formalism in the horizontal plane is the same), 
$\delta x'$ and $\delta y'$
expressed in eqs. \ref{eq:1p}, \ref{eq:1xx} and \ref{eq:111a},
respectively, into Taylor series:
$$\delta_y'={N_er_e\over \gamma_*}(
{1\over 2\sigma^2}y-{1\over 16\sigma^4}y^3+{1\over 192\sigma^6}y^5
-{1\over 3072\sigma^8}y^7$$
\begin{equation}
+{1\over 61440\sigma^{10}}y^9-{1\over 1474560\sigma^{12}}y^{11}
+{1\over 41287680\sigma^{14}}y^{13}
-\cdot \cdot \cdot )\qquad ({\rm RB})
\label{eq:2}
\end{equation}
$$\delta_x'=-{N_er_e\over 2\gamma_*}(
{2\over \sigma_x^2}x-{1\over 3\sigma_x^4}x^3+{1\over 30
\sigma_x^6}x^5
-{1\over 420\sigma_x^8}x^7$$
\begin{equation}
+{1\over 7560
\sigma_x^{10}}y^9-{1\over 166320
\sigma_x^{12}}x^{11}+{1\over 4324320
\sigma_x^{14}}x^{13}
-\cdot \cdot \cdot )\qquad ({\rm FB})
\label{eq:2x}
\end{equation}
$$\delta_y'=-{N_er_e\over \sqrt{2}\gamma_*}(
{2\over \sigma_x\sigma_y}y-{1\over 3\sigma_x\sigma_y^3}y^3+{1\over 20
\sigma_x\sigma_y^5}y^5
-{1\over 168\sigma_x\sigma_y^7}y^7$$
\begin{equation}
+{1\over 1728
\sigma_x\sigma_y^9}y^9-{1\over 21120
\sigma_x\sigma_y^{11}}y^{11}+{1\over 299520
\sigma_x\sigma_y^{13}}y^{13}
-\cdot \cdot \cdot )\qquad ({\rm FB})
\label{eq:2a}
\end{equation}
The differential equations of the motion of the test particle
in the transverse planes can be expressed as: 
$${d^2y\over ds^2}+K_y(s)y=-{N_er_e\over \gamma_*}(
{1\over 2\sigma^2}y-{1\over 16\sigma^4}y^3+{1\over 192\sigma^6}y^5$$
$$-{1\over 3072\sigma^8}y^7+{1\over 61440\sigma^{10}}y^9-{1\over 1474560\sigma^{12}}y^{11}$$
\begin{equation}
+{1\over 41287680\sigma^{14}}y^{13}-\cdot \cdot \cdot )\sum_{k=-\infty}^{\infty}\delta(s-kL)\qquad ({\rm RB})
\label{eq:3}
\end{equation}
$${d^2x\over ds^2}+K_x(s)x=-{N_er_e\over 2\gamma_*}(
{2\over \sigma_x^2}x-{1\over 3\sigma_x^4}x^3+{1\over 30
\sigma_x^6}x^5$$
$$
-{1\over 420\sigma_x8}x^7+{1\over 7560
\sigma_x^{10}}x^9-{1\over 166320
\sigma_x^{12}}x^{11}$$
\begin{equation}
+{1\over 4324320
\sigma_x^{14}}x^{13}-\cdot \cdot \cdot )\sum_{k=-\infty}^{\infty}\delta(s-kL)\qquad ({\rm FB})
\label{eq:3x}
\end{equation}
$${d^2y\over ds^2}+K_y(s)y=-{N_er_e\over \sqrt{2}\gamma_*}(
{2\over \sigma_x\sigma_y}y-{1\over 3\sigma_x\sigma_y^3}y^3+{1\over 20
\sigma_x\sigma_y^5}y^5$$
$$
-{1\over 168\sigma_x\sigma_y^7}y^7+{1\over 1728
\sigma_x\sigma_y^9}y^9-{1\over 21120
\sigma_x\sigma_y^{11}}y^{11}$$
\begin{equation}
+{1\over 299520
\sigma_x\sigma_y^{13}}y^{13}-\cdot \cdot \cdot )\sum_{k=-\infty}^{\infty}\delta(s-kL)\qquad ({\rm FB})
\label{eq:3a}
\end{equation}
where $K_x(s)$ and  
$K_y(s)$ describe the linear focusing of the lattice in the 
horizontal and vertical planes.
The corresponding Hamiltonians are expressed as:
$$H={p_y^2\over 2}+{K_y(s)\over 2}y^2+{N_er_e\over \gamma_*}
({1\over 4\sigma^2}y^2-{1\over 64\sigma^4}y^4+{1\over 1152\sigma^6}y^6$$
\begin{equation}
-{1\over 24576\sigma^8}y^8
+\cdot \cdot \cdot )\sum_{k=-\infty}^{\infty}\delta(s-kL)\qquad ({\rm RB}) 
\label{eq:4}
\end{equation}
$$H_x={p_x^2\over 2}+{K_x(s)\over 2}x^2+{N_er_e\over 2\gamma_*}(
{1\over \sigma_x^2}x^2-{1\over 12\sigma_x^4}x^4+{1\over 180
\sigma_x^6}x^6$$
\begin{equation}
-{1\over 3360\sigma_x^8}x^8
+\cdot \cdot \cdot )\sum_{k=-\infty}^{\infty}\delta(s-kL)\qquad ({\rm FB})
\label{eq:4x}
\end{equation}
$$H_y={p_y^2\over 2}+{K_y(s)\over 2}y^2+{N_er_e\over \sqrt{2}\gamma_*}(
{1\over \sigma_x\sigma_y}y^2-{1\over 12\sigma_x\sigma_y^3}y^4+{1\over 120
\sigma_x\sigma_y^5}y^6$$
\begin{equation}
-{1\over 1344\sigma_x\sigma_y^7}y^8
+\cdot \cdot \cdot )\sum_{k=-\infty}^{\infty}\delta(s-kL)\qquad ({\rm FB})
\label{eq:4a}
\end{equation}
where $p_x=dx/ds$ and $p_y=dy/ds$.
\par
\section{Review of the general analytical formulae for dynamic apertures} 
In ref. \cite{1} we have studied analytically the one dimensional ($y=0$) dynamic aperture of a storage ring
described by
the following Hamiltonian: 
$$
H={p^2\over 2}+{K(s)\over 2}x^2+{1\over 3!B\rho}{\partial^2B_z\over 
\partial x^2}x^3L\sum_{k=-\infty}^{\infty}\delta (s-kL)$$
\begin{equation}
+{1\over 4!B\rho}{\partial^3B_z\over
\partial x^3}x^4L\sum_{k=-\infty}^{\infty}\delta (s-kL)+\cdot \cdot \cdot
\label{eq:5}
\end{equation}
where
\begin{equation}
B_z=B_0(1+xb_1+x^2b_2+x^3b_3+x^4b_4+\cdot \cdot \cdot+x^{m-1}b_{m-1}+
\cdot \cdot \cdot)
\label{eq:6}
\end{equation}
The dynamic aperture corresponding to each multipole
is given as:
\begin{equation}
A_{dyna,2m,x}(s)
=\sqrt{2\beta_x(s)}\left({1\over m\beta_x^m(s_{2m})}\right)^{1\over 2(m-2)}
\left({\rho \over \vert b_{m-1}\vert L}
\right)^{1/(m-2)}
\label{eq:8}
\end{equation}
where $s_{2m}$ is the location of the $2m$th multipole, $\beta_x(s)$ is
the beta function in $x$ plane. Since these results
are general, we have tried to avoid  assigning  the freedom of motion,
$x$, a specific name,
such as horizontal, or vertical plane.
\par
\section{Beam-beam limited dynamic apertures}
To make use of the general dynamic aperture formulae recalled in section 3,
one needs only to find the equivalence relations by comparing three Hamiltonians
expressed in eqs. \ref{eq:4}, \ref{eq:4x}, and \ref{eq:4a} with eq. \ref{eq:5}, and it is found by analogy 
that:
\begin{equation}
{b_{m-1}\over \rho}L={N_er_e\over C_{m,RB}\gamma_* \sigma^m}\qquad ({\rm RB})
\label{eq:11}
\end{equation}
\begin{equation}
{b_{m-1}\over \rho}L={N_er_e\over C_{m,FB,x}2\gamma_* \sigma_x^m}\qquad ({\rm FB},x)
\label{eq:11a}
\end{equation}
\begin{equation}
{b_{m-1}\over \rho}L={N_er_e\over C_{m,FB,y}\sqrt{2}\gamma_* 
\sigma_x\sigma_y^{m-1}}\qquad ({\rm FB},y)
\label{eq:13a3}
\end{equation}
where $C_{m,RB}$, $C_{m,FB,x}$, and $C_{m,FB,y}$ are given in Table 1.
\begin{table}[t]
\begin{center}
\begin{tabular}{|l|l|l|l|l|l|l|}
\hline
m&4&6&8&10&12&14\\
\hline
$C_{m,RB}$&16&192&3072&61440&1474560&41287680\\
\hline
$C_{m,FB,x}$&3&30&420&7560&166320&4324320\\
\hline
$C_{m,FB,y}$&3&20&168&1728&21120&299520\\
\hline
\end{tabular}
\end{center}
\caption{summary of multipole coefficients}
\label{tab:1}
\end{table}  
Now by inserting eqs. \ref{eq:11}-\ref{eq:13a3} into
eq. \ref{eq:8} 
one 
can calculate dynamic apertures of different multipoles 
due to nonlinear beam-beam forces.
For example, one can get the dynamic apertures due to the beam-beam
octupole nonlinear force:
$$
A_{dyna,8,y}(s)={\sqrt{\beta_y(s)}\over \beta_y(s_{IP})}
\sqrt{\rho \over \vert b_3\vert L}$$
\begin{equation}
={\sqrt{\beta_y(s)}\over \beta_y(s_{IP})}\left({16\gamma_* \sigma^4\over N_er_e
}\right)^{1/2}\qquad ({\rm RB}) 
\label{eq:14}
\end{equation}
$$
A_{dyna,8,x}(s)={\sqrt{\beta_x(s)}\over \beta_x(s_{IP})}
\sqrt{\rho \over \vert b_3\vert L}$$
\begin{equation}
={\sqrt{\beta_x(s)}\over \beta_x(s_{IP})}\left({6\gamma_* 
\sigma_x^4\over N_er_e
}\right)^{1/2}\qquad ({\rm FB}) 
\label{eq:14x}
\end{equation}
$$
A_{dyna,8,y}(s)={\sqrt{\beta_y(s)}\over \beta_y(s_{IP})}
\sqrt{\rho \over \vert b_3\vert L}$$
\begin{equation}
={\sqrt{\beta_y(s)}\over \beta_y(s_{IP})}\left({3\sqrt{2}\gamma_* 
\sigma_x\sigma_y^3\over N_er_e
}\right)^{1/2}\qquad ({\rm FB}) 
\label{eq:14a}
\end{equation}
where $s_{IP}$ is the IP position.
Given the dynamic
aperture of the ring without the beam-beam effect as $A_{x,y}$, 
the total dynamic aperture including the beam-beam effect can be 
estimated usually as:     
\begin{equation}
A_{total,x,y}(s)={1\over \sqrt{{1\over A_{x,y}(s)^2}+{1\over A_{bb,x,y}(s)^2}}}
\label{eq:18}
\end{equation}
In the following we will consider the case of  
$A_{total,x,y}(s)\approx A_{bb,x,y}(s)$.
If we measure the beam-beam interaction limited dynamic apertures by the
beam sizes (the normalized beam-beam limited dynamic aperture), one gets:
\begin{equation}
{\cal R}_{y,8}={A_{dyna,8,y}(s)\over \sigma_{*}(s)}
=\left({16\gamma_* \sigma^2\over N_er_e\beta_y(s_{IP})}\right)^{1/2}
\qquad ({\rm RB})  
\label{eq:16}
\end{equation}
\begin{equation}
{\cal R}_{x,8}={A_{dyna,8,x}(s)\over \sigma_{*,x}(s)}
=\left({6\gamma_* \sigma_x^2\over N_er_e
\beta_x(s_{IP})}\right)^{1/2}
\qquad ({\rm FB})  
\label{eq:16x}
\end{equation}
\begin{equation}
{\cal R}_{y,8}={A_{dyna,8,y}(s)\over \sigma_{*,y}(s)}
=\left({3\sqrt{2}\gamma_* \sigma_x\sigma_y\over N_er_e
\beta_y(s_{IP})}\right)^{1/2}
\qquad ({\rm FB})  
\label{eq:16a}
\end{equation}
Recalling and using 
the definitions of the beam-beam tune shifts 
$\xi_x$ and $\xi_y$ in eqs. \ref{eq:22x} and \ref{eq:22}:
\begin{equation}
\xi_x^*={N_er_e\beta_{x,IP}\over 2\pi \gamma^*\sigma_x(\sigma_x+\sigma_y)}
\label{eq:22x}
\end{equation} 
\begin{equation}
\xi_y^*={N_er_e\beta_{y,IP}\over 2\pi \gamma^*\sigma_y(\sigma_x+\sigma_y)}
\label{eq:22}
\end{equation} 
one can simplify the above
defined normalized dynamic apertures. As general results one finds:
\begin{equation}
{\cal R}_{y,2m}={A_{dyna,2m,y}(s)\over \sigma_{*,y}(s)}
=\left({2^{m-2\over 2}C_{m,RB}\over 4\pi \sqrt{m}\xi_y^*}\right)^{1\over m-2}
\qquad ({\rm RB})  
\label{eq:16b}
\end{equation}
\begin{equation}
{\cal R}_{x,2m}={A_{dyna,2m,x}(s)\over \sigma_{*,x}(s)}
=\left({2^{m-2\over 2}C_{m,FB,x}\over 2\sqrt{m}\pi \xi_x^*}\right)^{1\over m-2}
\qquad ({\rm FB})  
\label{eq:16abx}
\end{equation}
\begin{equation}
{\cal R}_{y,2m}={A_{dyna,2m,y}(s)\over \sigma_{*,y}(s)}
=\left({2^{m-2\over 2}C_{m,FB,y}\over \sqrt{2m}\pi \xi_y^*}\right)^{1\over m-2}
\qquad ({\rm FB})  
\label{eq:16ab}
\end{equation}
Obviously, {\it the normalized beam-beam effect limited dynamic apertures
are determined only by the beam-beam tune shifts}. The impact of this
discovery will be more appreciated later. 
When the higher order multipoles effects ($2m >8$) can be neglected 
eqs. \ref{eq:14}, \ref{eq:14x} and \ref{eq:14a} 
give very good approximations 
dynamic apertures limited by one beam-beam IP. If there are $N_{IP}$ interaction points in a ring the dynamic apertures described in eqs. \ref{eq:14} and
\ref{eq:14a} will be reduced by a factor of $\sqrt{N_{IP}}$ (if these $N_{IP}$
interaction points can be regarded as independent). 
\par
\section{Beam lifetime due to beam-beam \\interactions}
We take the beam-beam limited dynamic
aperture as the rigid mechanical boundary, i.e., those particles which walk beyond this virtual boundary
will be regarded lost instantaneously. 
Based on this physical point of view we can calculate the beam lifetime due to 
the nonlinear beam-beam effect:
\begin{equation}
\tau_{bb}=
={\tau_y \over 2}\left({\sigma_y(s)^2 \over A_{dyna,y}(s)^2}\right)\exp\left(
{A_{dyna,y}(s)^2\over \sigma_y(s)^2}\right) 
\label{eq:19}
\end{equation}   
where $\tau_y$ is the synchrotron radiation damping 
time in vertical plane. It is interesting to note that 
eq. \ref{eq:19} is similar to 
but different from the quantum lifetime forluma \cite{3} where that
$\sigma_y(s)^2$ is used instead of $2\sigma_y(s)^2$. The reason is that the 
quantum radiation results in energy fluctuations of an electron and
$2\sigma_y(s)^2$ corresponds to the average of the square of the 
oscillation amplitude, however, the dynamic apertures calculated
above due to nonlinear forces are relevant only to the projected motions.
When the beam-beam octupole nonlinear force dominates the dynamic aperture,  
by inserting eqs. \ref{eq:16}, \ref{eq:16x}, and \ref{eq:16a} into eq. \ref{eq:19}, 
or inserting eqs. \ref{eq:16b}, \ref{eq:16abx}, and \ref{eq:16ab} into eq. \ref{eq:19} one gets:
\begin{equation}
\tau_{bb,y}^*={\tau_y^*\over 2}
\left({16\gamma_* \sigma^2\over N_er_e\beta_y(s_{IP})}\right)^{-1 } 
\exp\left({16\gamma_* \sigma^2\over N_er_e\beta_y(s_{IP})}\right)
\qquad ({\rm RB})
\label{eq:21}
\end{equation}  
\begin{equation}
\tau_{bb,x}^*={\tau_x^*\over 2}
\left({6\gamma_* \sigma_x^2\over N_er_e
\beta_x(s_{IP})}\right)^{-1 } 
\exp\left({6\gamma_* \sigma_x^2\over N_er_e
\beta_x(s_{IP})}\right)
\qquad ({\rm FB})
\label{eq:21x}
\end{equation}  
\begin{equation}
\tau_{bb,y}^*={\tau_y^*\over 2}
\left({3\sqrt{2}\gamma_* \sigma_x\sigma_y\over N_er_e
\beta_y(s_{IP})}\right)^{-1 } 
\exp\left({3\sqrt{2}\gamma_* \sigma_x\sigma_y\over N_er_e
\beta_y(s_{IP})}\right)
\qquad ({\rm FB})
\label{eq:21a}
\end{equation}  
or
\begin{equation}
\tau_{bb,y}^*={\tau_y^*\over 2}
\left({4\over \pi \xi_y^*}\right)^{-1 } 
\exp\left({4\over \pi \xi_y^*}\right)
\qquad ({\rm RB})
\label{eq:23}
\end{equation}
 \begin{equation}
\tau_{bb,x}^*={\tau_x^*\over 2}
\left({3\over \pi \xi_x^*}\right)^{-1 } 
\exp\left({3\over \pi \xi_x^*}\right)
\qquad ({\rm FB})
\label{eq:23x}
\end{equation}    
\begin{equation}
\tau_{bb,y}^*={\tau_y^*\over 2}
\left({3\over \sqrt{2}\pi \xi_y^*}\right)^{-1 } 
\exp\left({3\over \sqrt{2}\pi \xi_y^*}\right)
\qquad ({\rm FB})
\label{eq:23p}
\end{equation}  
More generally, one has:
\begin{equation}
\tau_{bb,2m,y}^*={\tau_y^*\over 2}
\left({2^{m-2\over 2}C_{m,RB}\over 4\pi \sqrt{m}\xi_y^*}\right)^{-{2\over m-2} } 
\exp\left(\left({2^{m-2\over 2}C_{m,RB}\over 4\pi \sqrt{m}\xi_y^*}\right)^{{2\over m-2} } \right)
\qquad ({\rm RB})
\label{eq:23q}
\end{equation}
\begin{equation}
\tau_{bb,2m,x}^*={\tau_x^*\over 2}
\left({2^{m-2\over 2}C_{m,FB,x}\over \pi 2\sqrt{m}\xi_y^*}\right)^{-{2\over m-2} }  
\exp\left(\left({2^{m-2\over 2}C_{m,FB,x}\over \pi 2\sqrt{m}\xi_x^*}
\right)^{{2\over m-2} } \right)
\qquad ({\rm FB})
\label{eq:23xxx}
\end{equation}     
\begin{equation}
\tau_{bb,2m,y}^*={\tau_y^*\over 2}
\left({2^{m-2\over 2}C_{m,FB,y}\over \pi \sqrt{2m}\xi_y^*}\right)^{-{2\over m-2} }  
\exp\left(\left({2^{m-2\over 2}C_{m,FB,y}\over \pi \sqrt{2m}\xi_y^*}\right)^{{2\over m-2} } \right)
\qquad ({\rm FB})
\label{eq:23a}
\end{equation}  
If we define the lifetime divided by the corresponding damping time as 
normalized beam lifetime, one finds that {\it the beam-beam effect limited
normalized lifetimes depend only on beam-beam tune shifts}.
Figs. \ref{fig5} and \ref{fig6} show the normalized beam lifetime with respect to the beam-beam tune shifts for both flat and round beams. 
\par
\section{The maximum beam-beam tune shifts for flat and round beams}
Now it is high time for us to discuss the maximum beam-beam tune shift problem.
In literatures the term ``maximum beam-beam tune shift'' of a specific 
machine is not well defined. One of the reasonable definitions 
would be that the maximum beam-beam tune shift corresponding to
a well defined minimum beam-beam limited lifetime. 
In this paper we propose to 
take this well defined minimum beam-beam limited lifetime as one hour
(the idea is to reduce eq. \ref{eq:18} to $A_{total}(s)\approx A_{bb}(s)$,
and to have a machine still working~!).
Assuming that 
for both round and flat beam cases one has the same $\tau_y$, from 
eqs. \ref{eq:23}, \ref{eq:23x} and \ref{eq:23p} one finds the following relations:
\begin{equation}
\xi_{y,max}^{RB}={4\sqrt{2}\over 3}\xi_{y,max}^{FB}=1.89\xi_{y,max}^{FB}
\label{eq:24}
\end{equation}  
and 
\begin{equation}
\xi_{x,max}^{FB}=\sqrt{2}\xi_{y,max}^{FB}
\label{eq:24x}
\end{equation}  
It is proved theoretically 
why round beam scheme can almost double the $\xi_{y,max}$ of flat beam scheme
as previously discovered in the numerical simulations \cite{7}\cite{8},
and why the vertical beam-beam tune shift reaches its limit earlier than the 
horizontal one.
Quantitatively, taking $\tau_y=30$ ms, one finds that $\xi_{y,max,FB}(\tau_{bb}=1$ 
hour)$=0.0447$, $\xi_{x,max,FB}(\tau_{bb}=1$ 
hour)$=0.0632$, and $\xi_{y,max,RB}(\tau_{bb}=1$ hour)$=0.0843$.
\par
Now we investigate how the order of nonlinear resonance affects the
maximum beam-beam tune shift. By using eqs. \ref{eq:23q}, \ref{eq:23x},
and \ref{eq:23a}, and assuming that $\tau_x=\tau_y$, one gets the maximum
beam-beam tune shift with respect to the order of nonlinear resonance, $m$,
as shown in Fig. \ref{fig9}, where each maximum beam-beam tune corresponds
to each dominating multipole resonance. For flat beams, it is obvious that
if the horizontal tune is not well chosen, the $\xi_{y,max}^{FB}$ can
be 0.032 instead of 0.0447, however, if the vertical resonances have been 
successfully avoided before $\xi_{x,FB}$ reaches its limit, one could
possibly obtain
$\xi_{x,max,FB}(\tau_{bb}=1$ hour)$=0.0632$ even difficultly.  
What should be stressed is that in choosing the working point in the
tune diagram, one has to pay attention to the 
nonlinear resonances of order as high
as 14. To explain qualitatively why the maximum beam-beam tune shifts
for both round and flat beams seem to be limited by the lowest order of 
resonance, i.e., the 1/4 resonance, we have plotted in Fig. \ref{fig10}
the sum of the
multipole strengths from m=4 to m=14 assuming that they have the same
strength, as expressed:
\begin{equation}
A(Q_y)=\sum_{m=4}^{m=14}(-1)^{m/2}\sin(2\pi m Q_y)
\label{eq:99}
\end{equation}  
On the same figure we have plotted also the first term (octupole)
in this summation with two opposite phases as compare references, and it is 
obvious that except two regions of $Q_y$, (0.2 to 0.3) and (0.7 to 0.8), one has
always that the amplitude of the sum is almost the same as that of the
octupole term, and in this case the dangerous $Q_y$ values are 0.225, 0.275, 0.725 and 0.775. 
Another reason for the lowest resonance dominating is that
the lower the resonance order the more stable the resonance facing to the
phase perturbations.  
\par 
Now we discuss briefly the choice of tunes (working point). Limited to 
one IP and the flat beam case, based on the original work 
of Bassetti (LNF-135, Frascati, Italy),
B. Richter has shown in ref. \cite{richt} that the tune should
be chosen just above an integer or half integer
to make a best use of dynamic beta effect, and this conclusion
has been experimentally observed in CESR \cite{Sagan}. 
Combining this information
with what suggested by eq. \ref{eq:99}, one concludes that the tune $Q$
should be chosen in the regions (0,0.2) or (0.5,0.7) to obtain
a maximum luminosity. 
If the collision is effectuated with a definite crossing angle some
important synchrobetatron nonlinear resonances, such as $3Q_x\pm Q_s=p$,
should be avoided also. More discussions on the crossing angle effects
will be given in section 8.
Taking CESR and PEP-II for examples, 
for CESR one finds $Q_x=10.523$ and $Q_y=9.597$ \cite{Sagan}, and for
PEP-II the actual operation
working points are $Q_x=0.610\pm 0.01$ and $Q_y=0.580\pm0.01$
for Low Energy Ring (LER) and $Q_x=0.565\pm 0.01$ and $Q_y=0.585\pm0.01$
for High Energy Ring (HER) \cite{placid}, which in principal consist
with our suggestion.     
\par
In this paper, under the assumption that the two colliding beams always
have the same transverse dimensions, we have arrived at the beam-beam 
effect determined lifetimes expressed in 
eqs. \ref{eq:23q}, \ref{eq:23xxx}, and \ref{eq:23a}. For a given
minimum normalized (with respect to the damping time) beam lifetime one gets
universal maximum beam-beam tune shift values corresponding to different
cases. In a real machine the situation can be more complicated, such as
the flip-flop phenomenon which breaks the symmetry assumed above, 
and in this case one can continue the discussion starting from
eqs. \ref{eq:14}, \ref{eq:14x}, and \ref{eq:14a} by differentiating
$\sigma_y$ from $\sigma_{*,y}$, and by replacing 
$\sigma$, $\sigma_x$ by $\Sigma/\sqrt{2}$, $\Sigma_x/\sqrt{2}$,
respectively, where $\Sigma=\sqrt{\sigma_*^2+\sigma^2}$ and 
$\Sigma_x=\sqrt{\sigma_{*,x}^2+\sigma_x^2}$.
We will not, however, 
continue our discussions in this direction in an exhaustive way.
\par
\section{Applications to some machines}
Let us look at three machines, PEP-II B-Factory \cite{Seeman}
and DA$\Phi$NE \cite{Vign}, and BTCF \cite{BTCF}, and the first two have been
put to operation. The relevant machine parameters are shown in Table 2.
Figs. \ref{fig1} and \ref{fig2} give the theoretical estimations for the 
beam-beam limited beam lifetimes in both PEP-II
LER and HER. 
Figs. \ref{fig7} and \ref{fig8} show the beam lifetimes versus
the beam-beam tune shifts in both LER and HER.
It is obvious that the nominal charge in the bunch of HER
is close to the limit which sets the beam lifetime in low energy
ring, however, the beam lifetime in HER is much longer than that in LER. 
The theoretical results consist with the experimental observation \cite{Seeman}.
Fig. \ref{fig3} shows the beam lifetime prediction for the DA$\Phi$NE
e$^+$e$^-$ collider with single IP. Finally, we study the beam-beam limited 
beam lifetime in BTCF (standard scheme) and 
the theoretical result is given in Fig. \ref{fig4} where the dot indicates
the designed beam lifetime.
\par
\begin{table}[t]
\begin{center}
\begin{tabular}{|l|l|l|l|l|l|l|l|}
\hline
Machine&$N_e$&$\beta_{y,IP}$ cm&$\sigma_{x,IP}$ $\mu$m&$\sigma_{y,IP}$ $\mu$m&$\gamma$&$\tau_y$ ms\\
\hline
PEP-II LER&6$\times 10^{10}$&1.5&157&4.78&6067&30\\
\hline
PEP-II HER&2.8$\times 10^{10}$&1.5&157&4.78&17613&18.3\\
\hline
DA$\Phi$NE&8.9$\times 10^{10}$&4.5&2100&21&998&35.7\\
\hline
BTCF&1.4$\times 10^{11}$&1&450&9&3914&31\\
\hline
\end{tabular}
\end{center}
\caption{Machine parameters}
\label{tab:2}
\end{table}
\par
\section{Discussion on the collision with a crossing angle}
To get a higher luminosity one could run 
a circular collider in the multibunch operation mode with a 
definite collision crossing angle. Different from the head-on collision
discussed above, the transverse kick received by a test particle due
to the space charge field of the counter rotating bunch will
depend on its longitudinal position with respect to the center of the bunch
which the test particle belongs to. 
In this section we consider first a flat beam colliding with another flat beam with a half crossing angle
of $\phi$ in the horizontal plane.
Due to the crossing angle the two curvilinear
coordinates of the
two colliding beams at the interaction point will be no longer coincide. 
The detailed discussion 
about the coordinates transformation can be found in ref. \cite{Hirata1}.
When the crossing angle is not too large one has:
\begin{equation}
x^*=x+z\phi
\label{eq:a1}
\end{equation}
where $x^*$ is the horizontal displacement of the test particle to 
the center of the colliding bunch, $z$ and $x$ are the longitudinal 
and horizontal displacements of the test particle
from the center of the bunch to which it belongs. 
Now we recall eq. \ref{eq:4x} which describes the
Hamiltonian of the horizontal motion of a test particle in the head-on collision mode, and by
inserting eq. \ref{eq:a1} into eq. \ref{eq:4x} we get:
$$H_x={p_x^2\over 2}+{K_x(s)\over 2}x^2+{N_er_e\over 2\gamma_*}(
{1\over \sigma_x^2}(x+z\phi )^2-{1\over 12\sigma_x^4}(x+z\phi)^4+{1\over 180
\sigma_x^6}(x+z\phi)^6$$
\begin{equation}
-{1\over 3360\sigma_x^8}(x+z\phi)^8
+\cdot \cdot \cdot )\sum_{k=-\infty}^{\infty}\delta(s-kL)\qquad ({\rm FB})
\label{eq:4xy}
\end{equation}
Since the test particle can occupy 
a definite $z$ within the bunch according to 
a certain probability distribution, say Gaussian,
it is reasonable to replace $z$ in
eq. \ref{eq:4xy} by $\sigma_z$, and in this way we reduce a two dimensional
Hamiltonian expressed in eq. \ref{eq:4xy} into a one dimensional one.
What should be noted is that eq. \ref{eq:4xy} takes only the 
test particle's longitudinal position into consideration which is regarded
as a small perturbation to the head-on collision case, and the geometrical
effect will included later.
To simplify our analysis we consider only
the lowest synchrobetatron nonlinear resonance, i.e., $3Q_x\pm Q_s=p$ (where
$Q_s$ is the synchrotron oscillation tune, and $p$ is an integer) which
turns out to be the most dangerous one \cite{Piw1}\cite{Piw2}.
Following the same procedure in section 4 one gets the dynamic aperture
due to the lowest synchrobetatron nonlinear resonance as follows:
\begin{equation}
A_{syn-beta,x}(s)=\left({2\beta_x(s)\over 3\beta_x(s_{IP})^3}\right)^{1/2}
{2\gamma_*\sigma_x^4\over N_er_e\sigma_z\phi}
\label{eq:a2}
\end{equation}
and 
\begin{equation}
{\cal R}_{syn-beta,x}={A_{syn-beta,x}(s)^2\over \sigma_x(s)^2}={2\over 3\pi^2}
\left({1\over \xi_x^*\Phi}\right)^2
\label{eq:a3}
\end{equation} 
where $\Phi ={\sigma_z\over \sigma_x}\phi$ is Piwinski angle.
Now we are facing a problem of how to combine the two effects:
the principal vertical beam-beam effect and the horizontal
crossing angle induced perturbation.
To solve this problem we assume that the total
beam lifetime due to the vertical and the horizontal crossing
angle beam-beam effects can be expressed as:
$$\tau_{bb,total}^*={\tau_x^*+\tau_y^*\over 4}
\left({1\over {1\over {\cal R}_{y,8,FB}}+{1\over {\cal R}_{syn-beta,x}}}\right)^{-1}\times$$
\begin{equation}
\exp\left({1\over {1\over {\cal R}_{y,8,FB}}+{1\over {\cal R}_{syn-beta,x}}}\right)
\qquad ({\rm FB})
\label{eq:a4}
\end{equation} 
where ${\cal R}_{y,8,FB}$ corresponds to eq. \ref{eq:16a}.  
After the necessary preparations, we can try to answer two frequently
asked questions. Firstly, for a machine working at the head-on collision 
beam-beam limit, how the beam lifetime depends on the crossing angle? 
Secondly, for a definite crossing angle, to keep the beam lifetime
the same as that of the head-on collision
at the beam-beam limit, how much one has to operate the machine 
below the designed head-on
peak luminosity?
To answer the first question we define
a lifetime reduction factor:
\begin{equation}
R(\Phi)={\tau_{bb,total}^*\over \tau_{bb,y}^*}
\qquad ({\rm FB})
\label{eq:a6}
\end{equation}      
where $\tau_{bb,y}^*$ is given in eq. \ref{eq:23p},
and $R(\Phi)$ will tell us to what extent one can increase $\Phi$.
Concerning the second question, one can
imagine to reduce the luminosity at beam-beam limit by a factor of 
$f(\Phi)$ in order to
against the additional lifetime reduction term due to 
the definite crossing angle. Physically, from eq. \ref{eq:a4} 
one requires: 
$$\left({A_{syn-beta,x}(s)^2\over \sigma_x(s)^2}\right)^{-1}
+\left({A_{dyna,crossing,8,y}(s)^2\over \sigma_y(s)^2}\right)^{-1}$$
\begin{equation}
=\left({A_{dyna,head-on,8,y}(s)^2\over \sigma_y(s)^2}\right)^{-1}
\qquad ({\rm FB})
\label{eq:a6a}
\end{equation}   
Mathematically, one has to solve the 
following equation to find the peak luminosity reduction factor $f(\Phi)$:
\begin{equation}
{3\pi^2 \xi_{x,design,FB}^2f(\Phi)^2\Phi^2\over 2}
+{\sqrt{2}\pi \xi_{y,max,FB}f(\Phi)\over 3}={\sqrt{2}\pi \xi_{y,max,FB}\over 3}
\qquad ({\rm FB})
\label{eq:a7}
\end{equation}   
\begin{equation}
f(\Phi)={-b_0+\sqrt{b_0^2+4a_0c_0}\over 2a_0}
\qquad ({\rm FB})
\label{eq:a8}
\end{equation} 
where $a_0=3\pi^2\xi_{x,design,FB}^2\Phi^2/2$,  
$b_0=c_0=\sqrt{2}\pi \xi_{y,max,FB}/3$, and $\xi_{x,max,FB}\approx 0.0447$. 
In fact, $f(\Phi)$ corresponds to the luminosity reduction due to the
synchrobetatron resonance, and to find out the total luminosity reduction
factor, one has to include the geometrical effects \cite{Piw0}\cite{Toge}. 
The total luminosity reduction factor can be expressed as follows:
\begin{equation}
F(\Phi)=f(\Phi)\left(1+\Phi^2\right)^{-1/2}
\qquad ({\rm FB})
\label{eq:a8a}
\end{equation}  
where hourglass effect is no taken into account (i.e. $\beta_{y,IP}>\sigma_z$).
Taking KEKB factory as an interesting example \cite{Seeman}, 
one has $\sigma_x=90$ $\mu$m,\linebreak  $\sigma_z=0.4$ cm, $\phi =11$ mrad, $\Phi =0.49$,
$\xi_{x,design,FB}=0.039$,
and by putting $\Phi =0.49$ into eq. \ref{eq:a8} one finds $F(0.49)=83.5\%$
which is very close to a three dimensional numerical simulation result, i.e., 
$85\%$ of designed head-on luminosity, given in ref. \cite{Ohmi}.  
In Figs. \ref{fig5} and \ref{fig6} we show how $R(\Phi)$ and $F(\Phi)$ depend
on Piwinski angle where $\xi_{x,design,FB}=0.039$ has been used in 
Fig. \ref{fig6}.
Finally, when the crossing angle is in the vertical plane or the beam is round,
one gets:
\begin{equation}
{\cal R}_{syn-beta,y}={1\over 3\pi^2}\left({r\over \xi_y^*\Phi}\right)^2
\qquad ({\rm FB})
\label{eq:a9}
\end{equation}
and  
\begin{equation}
{\cal R}_{syn-beta,y}={32\over 27\pi^2}\left({1\over \xi_y^*\Phi}\right)^2
\qquad ({\rm RB})
\label{eq:a10}
\end{equation} 
where $r=\sigma_y/\sigma_x$ and $\Phi ={\sigma_z\over \sigma_x}\phi$ as
defined before. Replacing ${\cal R}_{syn-beta,x}$ in eq. \ref{eq:a4}
by eq. \ref{eq:a9} or eq. \ref{eq:a10} and following the same procedure
shown above 
one can easily make the corresponding discussions on the luminosity 
reduction effects. 
What should be remembered is that the geometrical luminosity
reduction factors for the vertical crossing angle and the round beam cases are
$(1+(\Phi/r)^2)^{-1/2}$ and $(1+\Phi^2)$, respectively.
\par

\section{Conclusion}
In this paper we have established 
analytical formulae for the beam-beam interaction 
limited dynamic apertures and beam lifetimes 
in e$^+$e$^-$ circular colliders for both round and flat beam cases. 
It is shown analytically why for flat colliding beams one has always $\xi_{y,max}$
around 0.045 and why this value can be almost doubled by using 
round beams. Applications to the
machines, such as PEP-II, DA$\Phi$NE, and BTCF have been made. Finally,
the luminosity reduction due to a crossing angle has been discussed,
and an analytical formula for the luminosity reduction factor
is derived and compared with a numerical simulation result for KEKB factory.

\par
\section{Acknowledgement}
The author thanks J. Ha\"\i ssinski for his careful reading of the manuscript,
critical comments, and drawing my attention to the paper of B. Richter.
Stimulating discussions with A. Tkatchenko and J. Le Duff are appreciated.

\newpage
\begin{figure}[t]
\vskip 0. true cm
\vspace{6.0cm}
\includegraphics{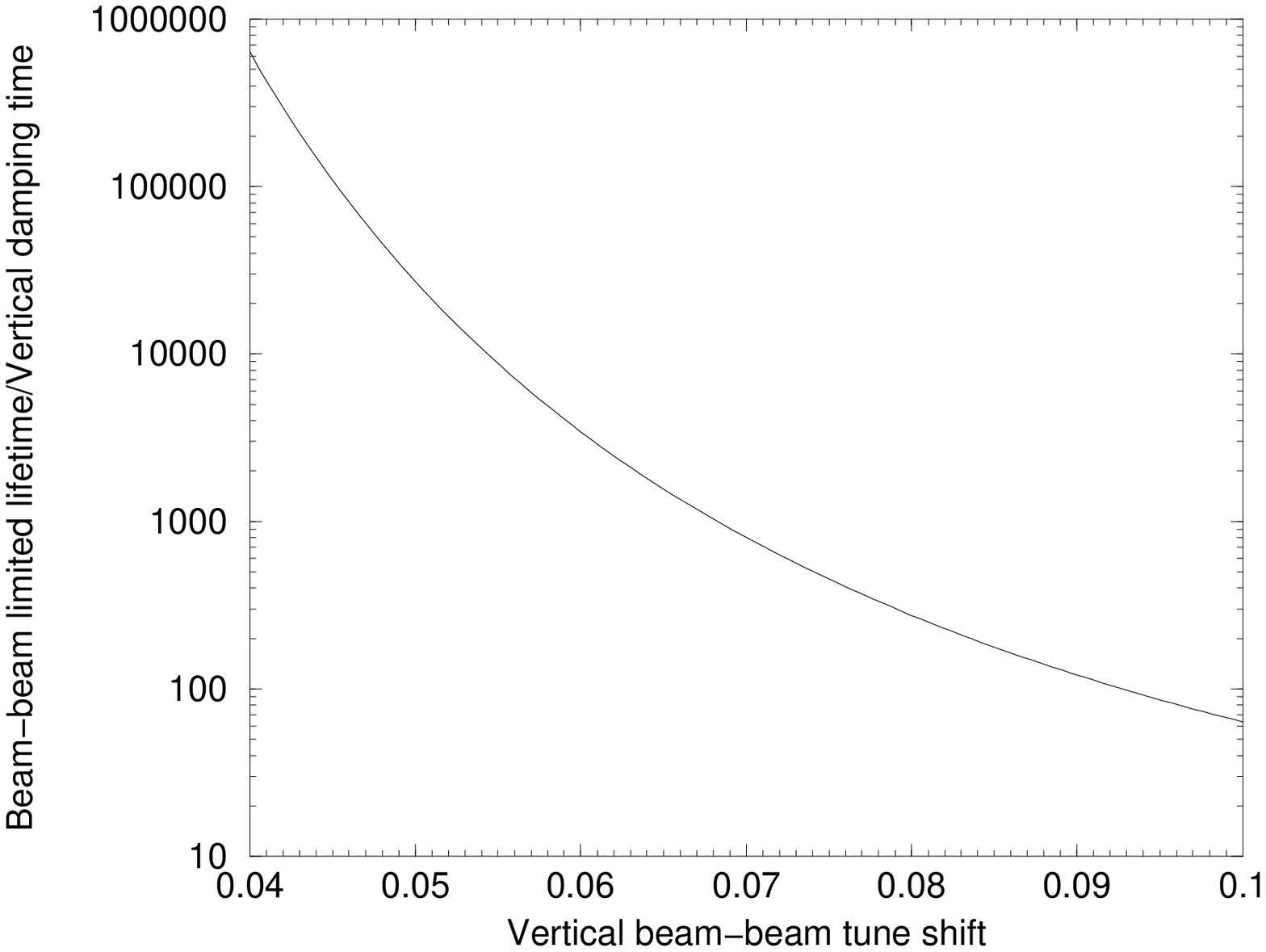}
\vskip -. true cm
\caption{${\tau_{bb}^*\over \tau_y^*}$ vs $\xi_y^*$ for flat beam case.
\label{fig5}}
\end{figure}
\begin{figure}[t]
\vskip 0.5 true cm
\vspace{6.0cm}
\includegraphics{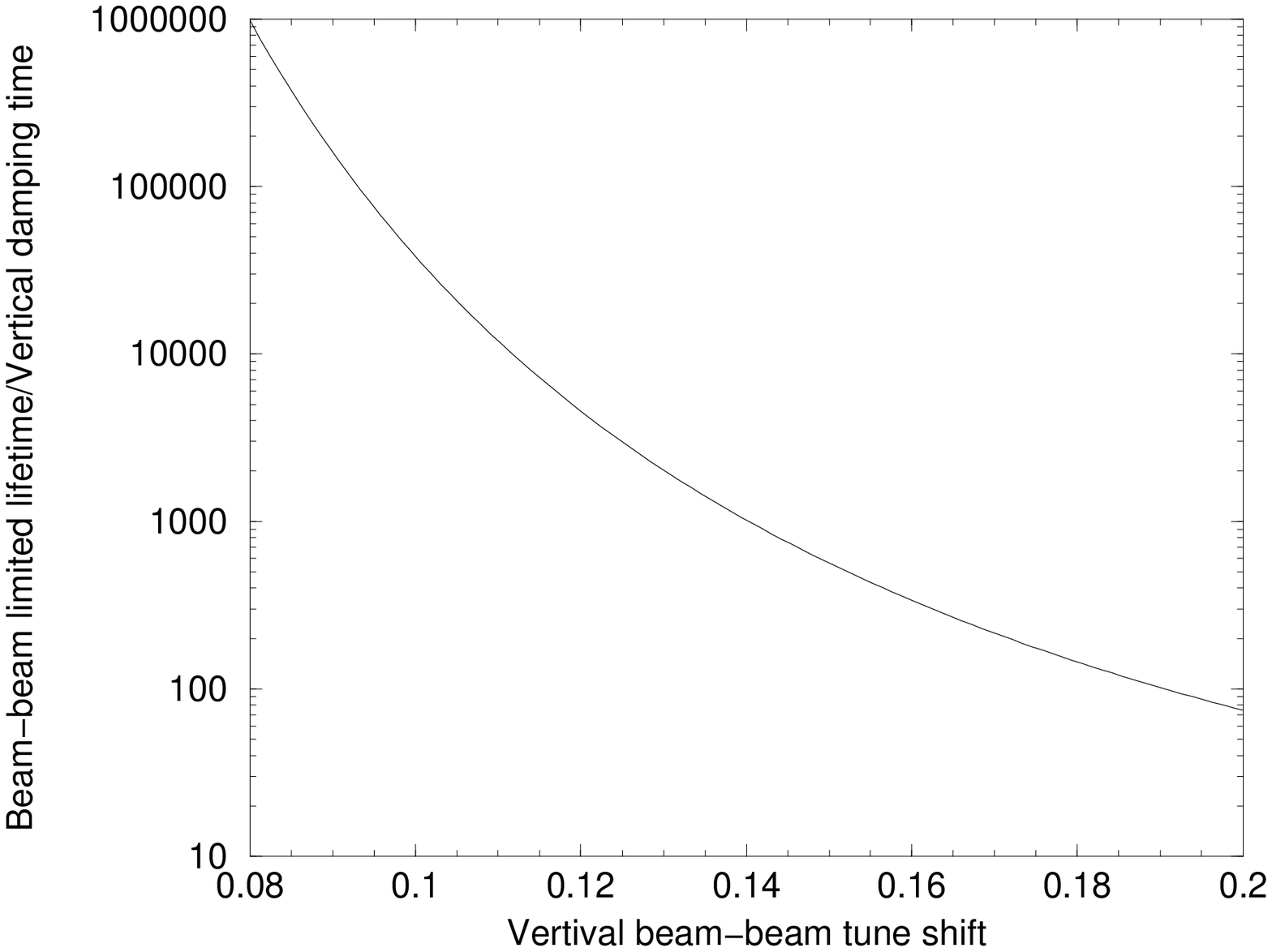}
\vskip -. true cm
\caption{${\tau_{bb}^*\over \tau_y^*}$ vs $\xi_y^*$ for round beam case.
\label{fig6}}
\end{figure}
\newpage
\begin{figure}[t]
\vskip 0. true cm
\vspace{6.0cm}
\includegraphics{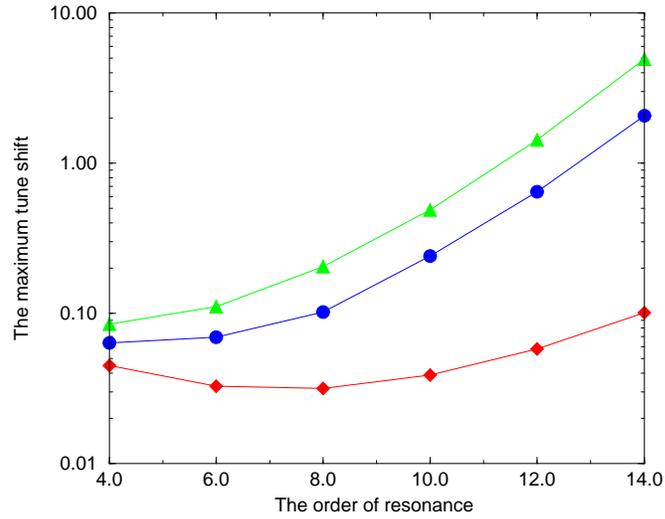}
\vskip -. true cm
\caption{The maximum beam-beam tune shift vs the resonance order. For a
flat beam the dotted line and the diamonded line correspond to the
horizontal and vertical tune resonances, respectively. The triangled line
corresponds to the vertical resonance for a round beam case.  
\label{fig9}}
\end{figure}
\begin{figure}[h]
\vskip 0. true cm
\vspace{6.0cm}
\includegraphics{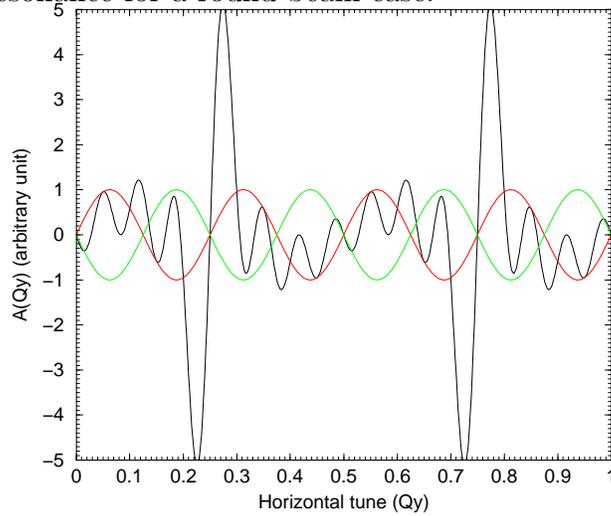}
\vskip -. true cm
\caption{The nonlinear perturbation amplitude vs the vertical tune. The
two inversely phased sinusoidal curves correspond to the amplitude of the
octupole term, and the fast oscillating curve is the sum 
of the multipoles of the same amplitude from octupole to 28 poles.  
\label{fig10}}
\end{figure}
\newpage
\begin{figure}[t]
\vskip 0. true cm
\vspace{6.0cm}
\includegraphics{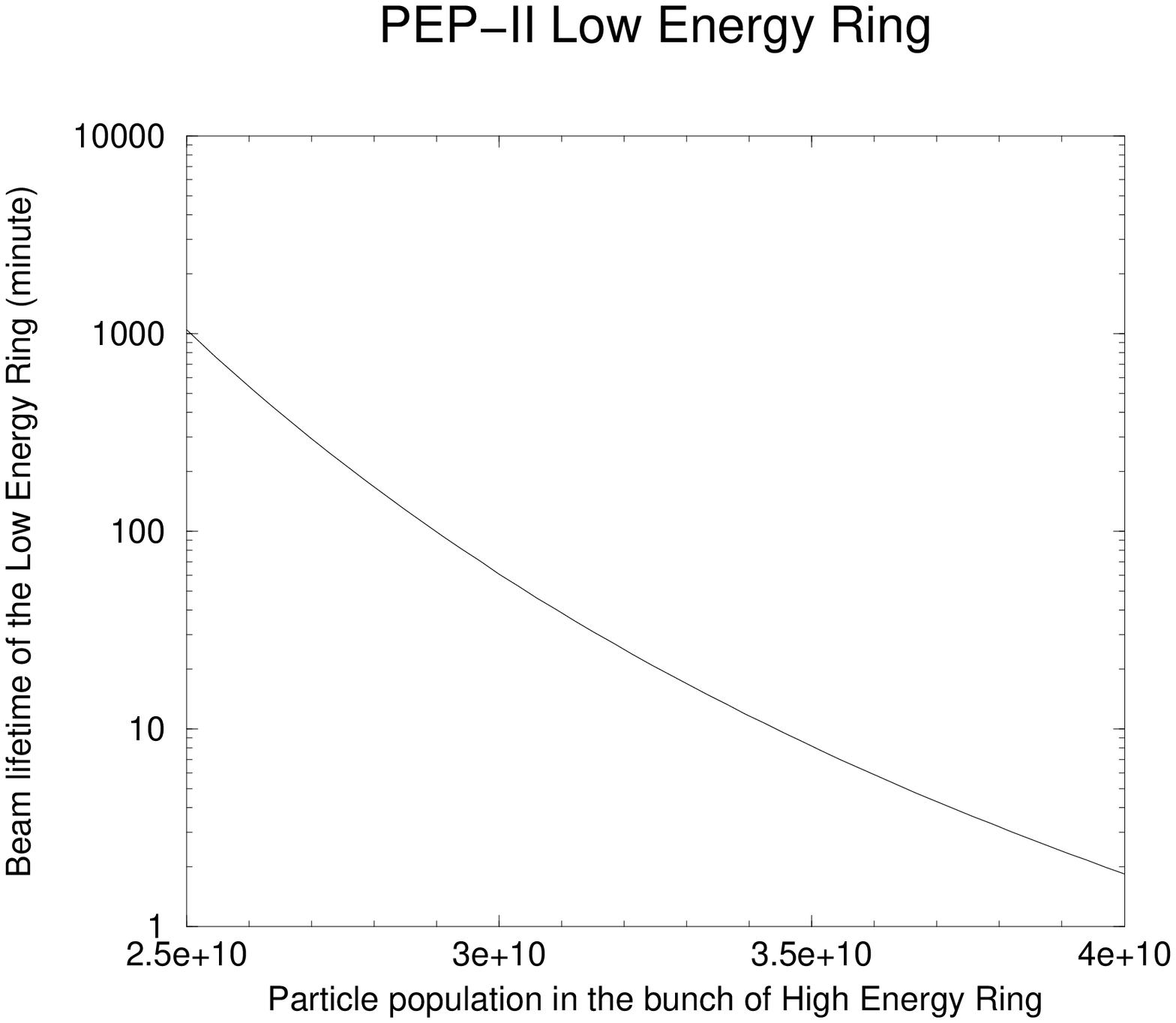}
\vskip -. true cm
\caption{The beam lifetime due to the
beam-beam interaction vs the particle population in the bunch
in the low energy ring of PEP-II.
\label{fig1}}
\end{figure}
\begin{figure}[h]
\vskip 0.5 true cm
\vspace{6.0cm}
\includegraphics{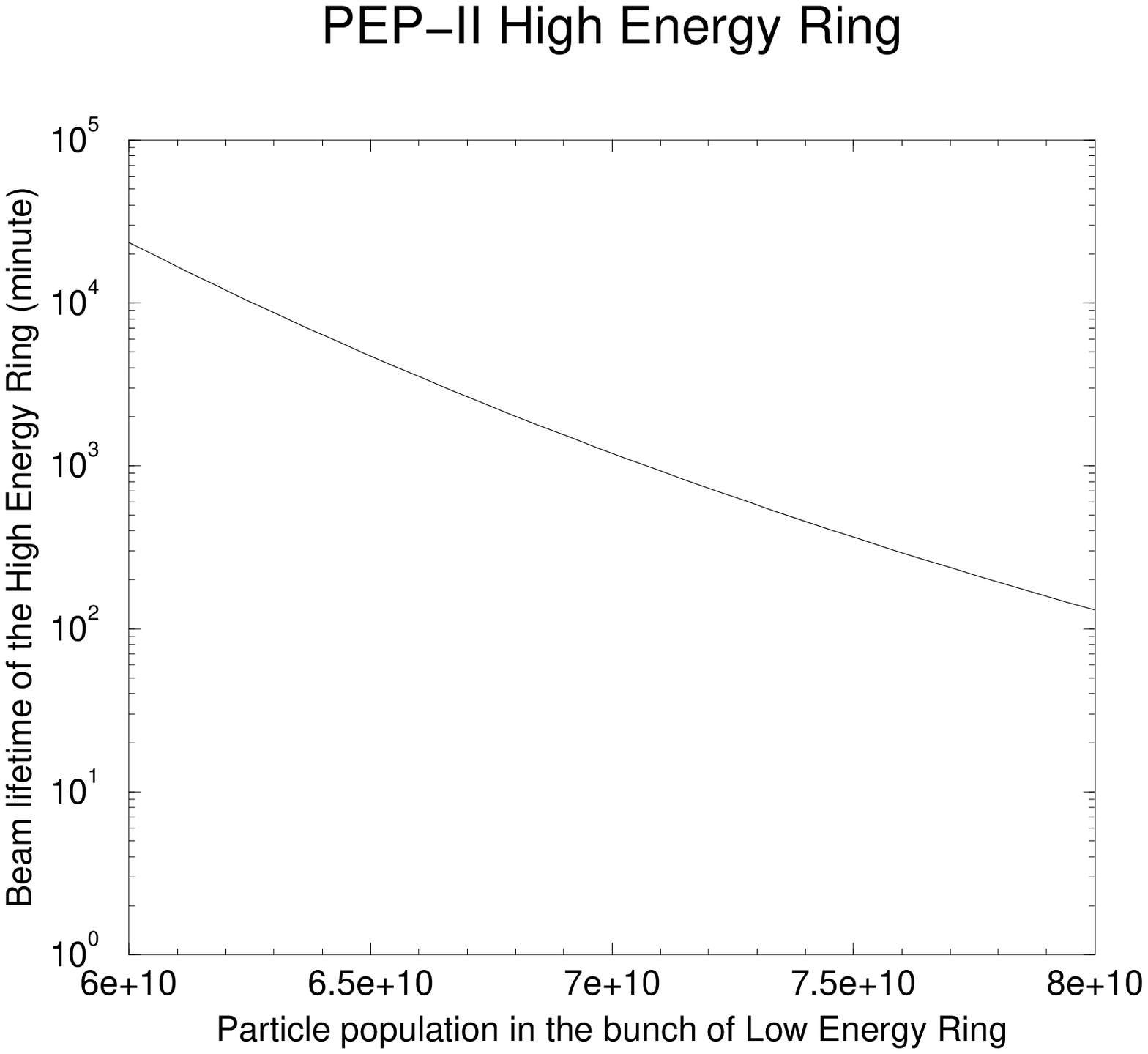}
\vskip -. true cm
\caption{The beam lifetime due to the
beam-beam interaction vs the particle population in the bunch in the high
energy ring of PEP-II.
\label{fig2}}
\end{figure}
\newpage
\begin{figure}[t]
\vskip 0. true cm
\vspace{6.0cm}
\includegraphics{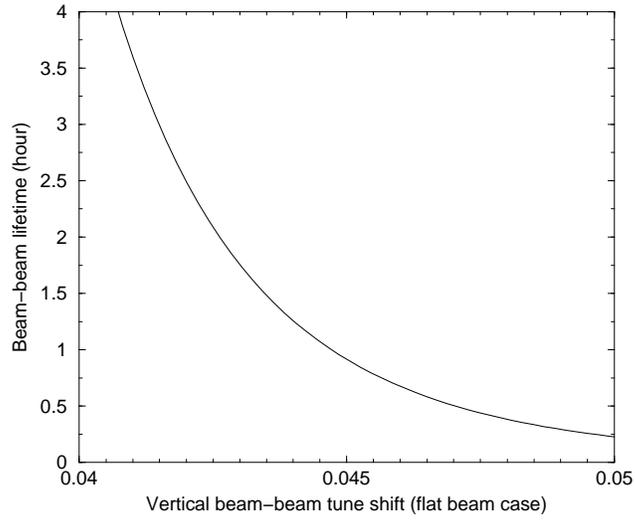}
\vskip -. true cm
\caption{PEP-II Low Energy Ring (flat beam): the beam-beam limited
lifetime $\tau_{bb}^*$ (hours) vs the vertical beam-beam tune shift $\xi_y$.
\label{fig7}}
\end{figure}
\begin{figure}[h]
\vskip 0.5 true cm
\vspace{6.0cm}
\includegraphics{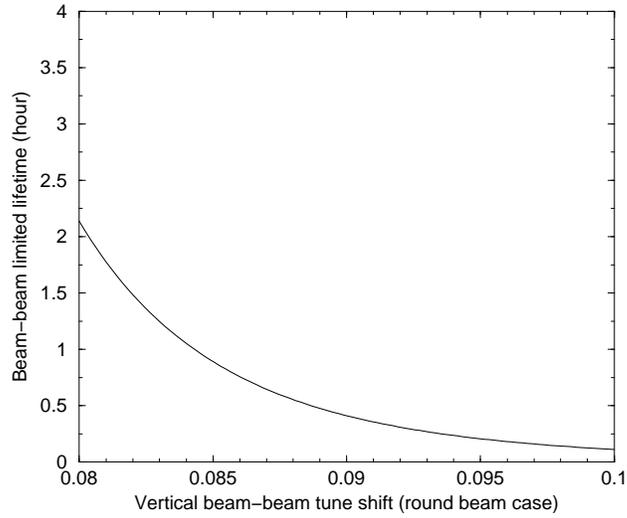}
\vskip -. true cm
\caption{PEP-II Low Energy Ring (round beam is assumed): the beam-beam limited
lifetime $\tau_{bb}^*$ (hours) vs the vertical beam-beam tune shift $\xi_y$.
\label{fig8}}
\end{figure}
\newpage
\begin{figure}[t]
\vskip 0. true cm
\vspace{6.0cm}
\includegraphics{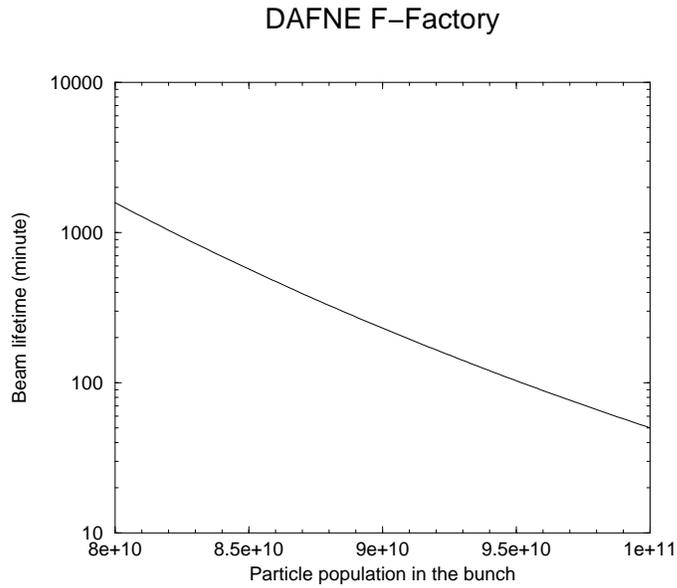}
\vskip -. true cm
\caption{The beam lifetime due to the
beam-beam interaction vs the particle population in the bunch
in the DA$\Phi$NE $\Phi$-Factory.
\label{fig3}}
\end{figure}
\begin{figure}[h]
\vskip 0.5 true cm
\vspace{6.0cm}
\includegraphics{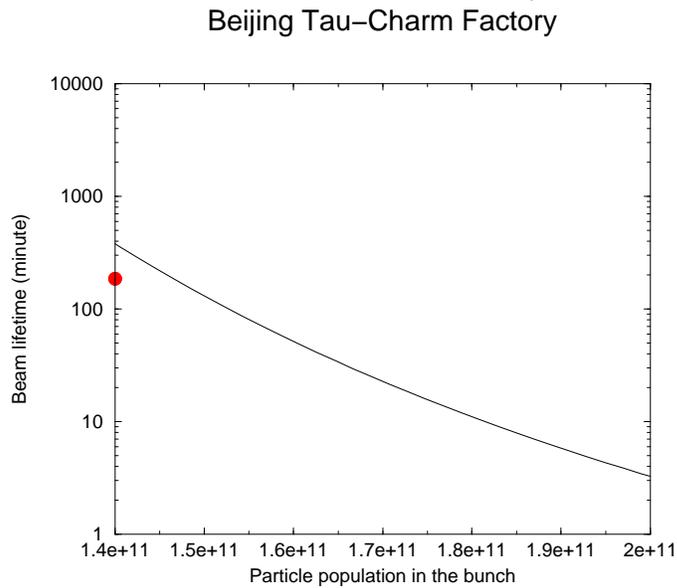}
\vskip -. true cm
\caption{The beam lifetime due to the
beam-beam interaction vs the particle population in the bunch in the 
Beijing $\tau$-C Factory (standard scheme).
\label{fig4}}
\end{figure}
\newpage
\begin{figure}[t]
\vskip 0. true cm
\vspace{6.0cm}
\includegraphics{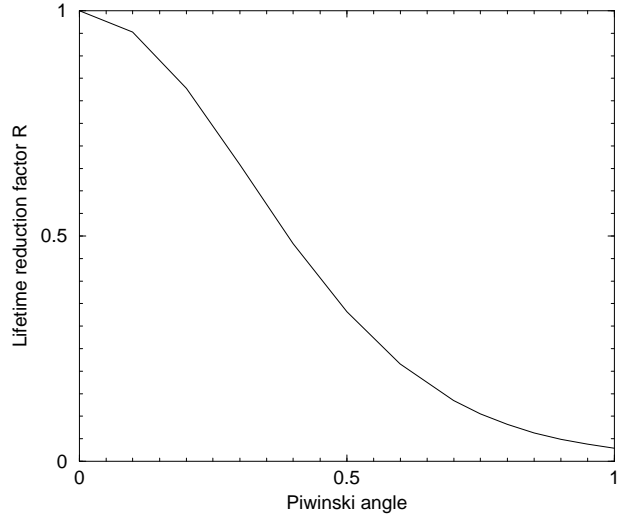}
\vskip -. true cm
\caption{The lifetime reduction factor $R(\Phi)$ vs Piwinski angle 
$\Phi$.
\label{fig5}}
\end{figure}
\begin{figure}[h]
\vskip 0.5 true cm
\vspace{6.0cm}
\includegraphics{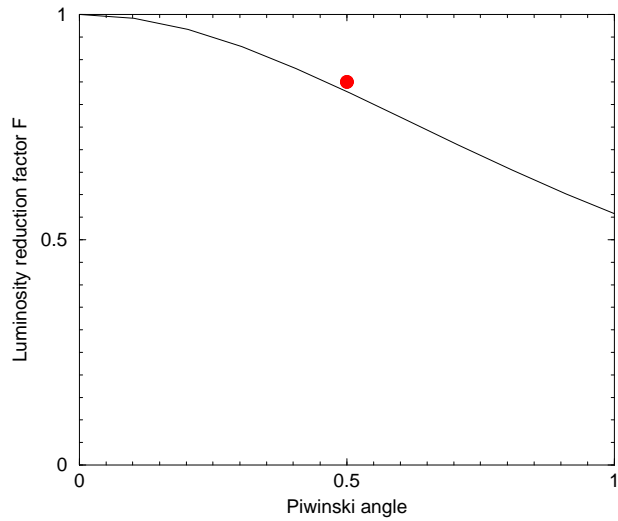}
\vskip -. true cm
\caption{The luminosity reduction factor $F(\Phi)$ vs Piwinski angle 
$\Phi$. The curve is obtained by taking $\xi_{x,design,FB}=0.039$ (KEKB),
and the dot is the numerical simulation result given in ref. \cite{Ohmi}.
\label{fig6}}
\end{figure}

\end{document}